\documentclass{article}      
\usepackage{graphicx}
\usepackage{epstopdf}
\title{The hypercentral Constituent Quark Model}  

\author{M.M. Giannini\\
Dipartimento di Fisica dell'Universit\`a di Genova\\
and \\
I.N.F.N., Sezione di Genova\\
E. Santopinto\\
I.N.F.N., Sezione di Genova\\
}

\date{}
\begin{document}

\maketitle  

\begin{abstract}
 The hypercentral Constituent Quark Model is presented and its application to 
the description of the electromagnetic properties of baryons is reviewed. The results 
concerning the elastic nucleon form factors and the electromagnetic excitation of baryon 
resonances are compared with the recent experimental data
\end{abstract}

\section{Introduction}

 The hypercentral Constituent Quark Model (CQM) has been proposed some years ago \cite{pl} and it has been applied to the description of non strange baryons.
 The hCQM contains, as it will be seen, only three free parameters, which are fitted to the  baryon spectrum. Once the parameters have been fixed, the model is completely determined and its results provide predictions for the baryons properties. In particular we shall consider the helicity amplitudes for the electromagnetic excitation of the baryon resonances and the elastic nucleon form factors. The goal is to describe data, if possible, but mainly to understand the  mechanisms underlying the hadron dynamics, with particular attention to what is missing and should be introduced in the model in order to improve the agreement with data.

\section{The Model}

The hCQM, as other ones proposed in the literature \cite{is,cq}, is based on the idea of Constituent Quarks \cite{morp65} (CQ). CQs are effective degrees of freedom, which describe the main quantum numbers of hadrons, but, at variance with QCD quarks, may acquire mass and even size.

\begin{table}
\begin{tabular}{lccccc}
\hline
$(56, 0^+)$ & $P_{11}(938)$ & $P_{33}(1232)$ &     & & \\
\hline
$(56*,0^+)$ & $P_{11}(1440)$  & $P_{33}(1600)$ &  & & \\
\hline
$(70, 1^-)$  & $D_{13}(1520)$ & $S_{11}(1535)$ & $S_{11}(1650) $  & $D_{13}(1700)$ &\\
 &  & $S_{31}(1620)$ & $D_{15}(1675)$ & $D_{33}(1700)$  &  \\
\hline
$(56, 2^+)$ & $F_{15}(1680)$ & $P_{13}(1720)$ &  $F_{35}(1910)$ & $P_{33}(1920)$  & $F_{37}(1950)$ \\
\hline
$(70, 0^+)$ & $P_{11}(1710)$ & $P_{31}(1910)$ & & &\\
\hline
\end{tabular}
\caption{The SU(6) multiplets of the lower  non strange baryon states}
\label{a}
\end{table}

It has been observed \cite{deru} that, according to Lattice QCD calculations \cite{wil}, the quark interaction contains a long range spin-independent confinement and a short range spin dependent term. The latter can be attributed to the one-gluon exchange interaction and treated as a perturbation. The dominance of the spin independent term allows to group the three-quark states into SU(6)-multiplets. The baryon states are given by the product of three quark states, each being given by a six-component spinor. The three quark states can be decomposed into irreducible representations of SU(6) according to the scheme:

\begin{equation}
6 \otimes 6 \otimes 6 = 20  \oplus  70  \oplus  70  \oplus  56
\end{equation}
where the representations have been labeled by their dimensions $d$. Introducing the total orbital angular momentum $L$ and the parity $\pi$, the SU(6) configurations can be labeled by $(d, L^\pi)$.In Table (\ref{a}) we show how the observed non strange baryon states can be arranged into SU(6) multiplets. 

The internal quark motion is described by the Jacobi coordinates
\begin{equation}
\vec{\rho}~=~ \frac{1}{\sqrt{2}} (\vec{r}_1 - \vec{r}_2) ~,
~~~~\vec{\lambda}~=~\frac{1}{\sqrt{6}} (\vec{r}_1 + \vec{r}_2 - 2\vec{r}_3) ~,
\end{equation}
\noindent or equivalently, $\rho$, $\Omega_{\rho}$, $\lambda$, 
$\Omega_{\lambda}$.  In order to describe the three-quark dynamics it is 
convenient to introduce the
hyperspherical coordinates, which are obtained substituting the absolute
values $\rho$ and $\lambda$ by
\begin{equation}
x=\sqrt{{\vec{\rho}}^2+{\vec{\lambda}}^2} ~~,~~ \quad
t=arctg(\frac{{\rho}}{{\lambda}}),
\end{equation}
where $x$ is the hyperradius and $t$ the hyperangle. In this way the angular-hyperangular part of the three quark wave functions is described by the hyperspherical harmonics (h.h.) \cite{baf} $Y_{[\gamma]}(\Omega)$, which are eigenstates of $L^2(\Omega)$, the Casimir operator of $O(6)$:
\begin{equation}
L^2(\Omega) ~Y_{[\gamma]}(\Omega)~=~-\gamma(\gamma+4) Y_{[\gamma]}(\Omega)
\end{equation}
where $\gamma=2n+l_{\lambda}+l_{\rho}$, n being a non zero integer, $l_{\lambda}$ and $l_{\rho}$ the quark orbital angular momenta associated with the respective Jacobi variables and $\Omega$ denotes the angular-hyperangular variables.

The quark interaction can be expanded in a h.h. series, the first term depending on the hyperradius $x$ only:
\begin{equation}
\Sigma_{i<j} ~V(r_{ij})~=~V(x) + .....
\end{equation}

In the hypercentral constituent quark model (hCQM), the quark potential is assumed to depend on the hyperradius $x$ only, that is to be 
hypercentral. Therefore, $V~=~V(x)$ is in general a three-body potential,
since the
hyperradius $x$ depends on the coordinates of all the three quarks. In the three-quark wave function one can then factor out the hyperangular part $Y_{[\gamma]}(\Omega)$. The remaining hyperradial part of the wave function ${\psi}_{\gamma,\nu}(x)$ is 
determined by the hypercentral Schr\"{o}dinger equation:
\begin{equation}
\frac{{d}^2}{dx^2}+\frac{5}{x}\frac{d}{dx}-\frac{\gamma(\gamma+4)}{x^2}
{\psi}_{\gamma,\nu}(x)=-2m[E-V(x)]{\psi}_{\gamma,\nu}(x),
\end{equation}\label{rad}
where  $v$ is a non negative integer number counting the number of nodes.

There are  two hypercentral potentials which lead to 
analytical solutions. First,
the h.o. potential, which has a two-body character, turns out to be exactly
hypercentral, since
\begin{equation}
\sum_{i<j}~\frac{1}{2}~k~(\vec{r_i} - \vec{r_j})^2~=~\frac{3}{2}~k~x^2~=
~V_{h.o.}(x).
\end{equation}
The second one is the 'hypercoulomb' potential
\begin{equation}\label{hcb}
V_{hyc}(x)= -\frac{\tau}{x}.
\end{equation}
This potential is not confining, however it is interesting \cite{bis} because
 it has an exact degeneracy between
the first $0^+$ excited state and the first $1^-$ states \cite{rich,hca,br},
which can be respectively identified with the Roper resonance and
the negative parity resonances. This degeneracy seems to be in agreement with
phenomenology and is typical of an underlying O(7) symmetry \cite{br}.
This feature cannot be reproduced in
models with only two-body forces and/or harmonic oscillator bases since the
excited $L = 0$ state, having one more node, lies above the $L =1 $ state 
\cite{rich}.

In the hCQM the quark potential is assumed in the 
form \cite{pl}
\begin{equation}\label{eq:pot}
V(x)= -\frac{\tau}{x}~+~\alpha x~~~~,
\end{equation}
\noindent 
that means a coulomb-like term plus a linear confining term as suggested by 
lattice QCD calculations \cite{bali}.
In order to describe the splittings within the $SU(6)$-multiplets 
we introduce a hyperfine interaction $H_{hyp}$ of the standard form \cite{is} and we treat
it as a perturbation. The three quark hamiltonian is given then
\begin{equation}\label{hCQM}
H~=~\frac{p_{\lambda}^2}{2m}~+~\frac{p_{\rho}^2}{2m}~ -\frac{\tau}{x}~+~\alpha x~+~H_{hyp}
\end{equation}

Having fixed the quark mass $m$ to $1/3$ of the nucleon mass, the 
remaining three free parameters ($\tau$, $\alpha$ and the strength of the 
hyperfine interaction) are fitted to the spectrum. The strength of the
hyperfine 
interaction is determined by the $\Delta$ - Nucleon mass difference and 
the remaining parameters are given by $\tau=4.59$
and $\alpha=1.61~fm^{-2}$ \cite{pl}.  

The confinement part of the interaction is not so effective in the low $x$ region, where the quark wave function is mainly concentrated and can therefore be treated perturbatively \cite{sig}; in this way, with a simplified form of the spin interaction, the model can be  be formulated in an analytical approach \cite{sig}.
The SU(6) violation can be given also by an isospin dependent term \cite{iso}, leading to a substantial improvement of the spectrum.

However in the following we shall use the interaction given by Eq. (\ref{hCQM}). Having fixed the parameters 
of the potential, the wave functions of the various 
resonances are 
completely determined and can be used for the prediction of various electromagnetic baryon properties.

\section{The helicity amplitudes}

The electromagnetic transition amplitudes, 
$A_{1/2}$, $A_{3/2}$ and $S_{1/2}$, are defined as the matrix elements of
the quark electromagnetic interaction, $A_{\mu} ÷J^{\mu}$, between the nucleon, 
$N$, and the resonance, $B$, states:
\begin{equation} \label{eq:hel}
\begin{array}{rcl}
A_{1/2}&=&   \sqrt{\frac{2 \pi \alpha}{k}}   \langle B, J', J'_{z}=\frac{1}{2}\ | J_{+}| N, J~=~
\frac{1}{2}, J_{z}= -\frac{1}{2}\
\rangle\\
& & \\
A_{3/2}&=& \sqrt{\frac{2 \pi \alpha}{k}}  \langle B, J', J'_{z}=\frac{3}{2}\ |J_{+} | N, J~=~
\frac{1}{2}, J_{z}= \frac{1}{2}\
\rangle\\
& & \\
S_{1/2}&=&   \sqrt{\frac{2 \pi \alpha}{k}}   \langle B, J', J'_{z}=\frac{1}{2}\ | J_{z}| N, J~=~
\frac{1}{2}, J_{z}= -\frac{1}{2}\
\rangle\\\end{array}
\end{equation}
$J_{\mu}$ is the electromagnetic current carried by quarks and will be used in its non relativistic  form 
\cite{cko, ki}; $k$ is the photon momentum in the Breit frame. 

\begin{figure}[h]
  \includegraphics[height=.4\textheight]{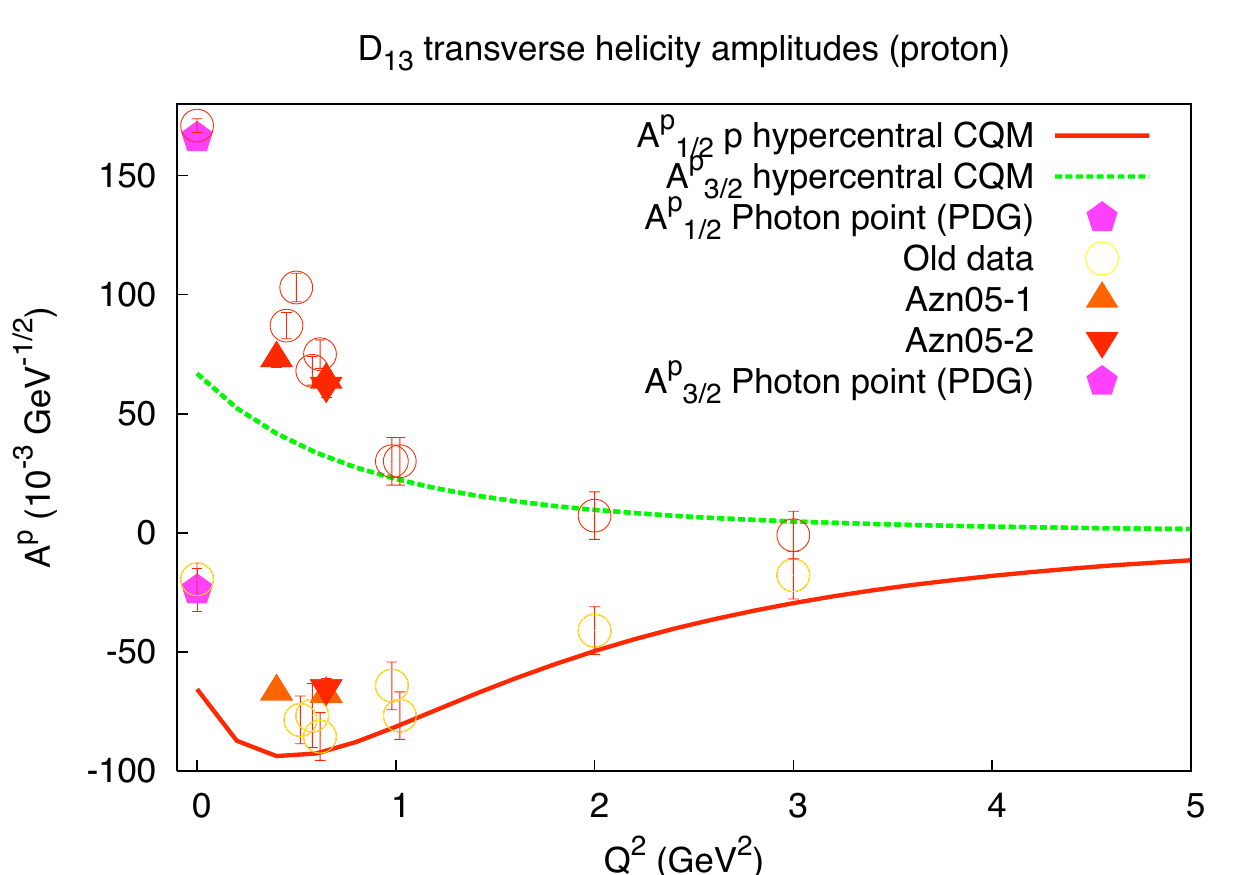}
  \caption{(Color online) The transverse helicity amplitudes for the D13(1520) resonance, calculated with the hCQM of Eq. (\ref{hCQM}). Data are from \cite{bu,azn05-1,azn05-2}. The photon point is also shown \cite{pdg}. }
\end{figure}

The results for the photocouplings, that is the transverse amplitudes with zero photon tetramomentum ($Q^2=0$), the results of the hCQM compare favourably with other model \cite{aie}.The overall trend is reproduced, but all models suffer of a lack of strength; the similarities among the various models can be ascribed to a common underlying SU(6) structure.

The three-quark wave functions calculated with the hCQM of Eq. (\ref{hCQM}) can be used for the prediction of the $Q^2$ behavior of the helicity amplitudes \cite{aie2}. The results for the transverse excitation to the negative parity resonances have been already published \cite{aie2}, while a systematic calculation of all excitations, both transverse and longitudinal will appear soon \cite{sg}.

As an example, the hCQM results for the D13(1520) and the S11(1535) resonances 
\cite{aie2}, are given in Fig. 1 and 2, respectively. The agreement in the case of the
S11 is remarkable, the more so since the hCQM curve has been published \cite{aie2} well in advance with respect to the recent TJNAF data. 
In general the $Q^2$ behaviour is reproduced, except for
discrepancies at small $Q^2$, especially in the
$A^{p}_{3/2}$ amplitude of the transition to the $D_{13}(1520)$ state. 
 The kinematical relativistic corrections at the level of
boosting the nucleon  and the resonances states to a common frame do not modify substantially the non-relativistic results  \cite{mds2}. Therefore these discrepancies can  be ascribed  to the lack of explicit quark-antiquark configurations \cite{aie2,es}, which are expected to be important at low $Q^{2}$ .

\begin{figure}[h]
  \includegraphics[height=.4\textheight]{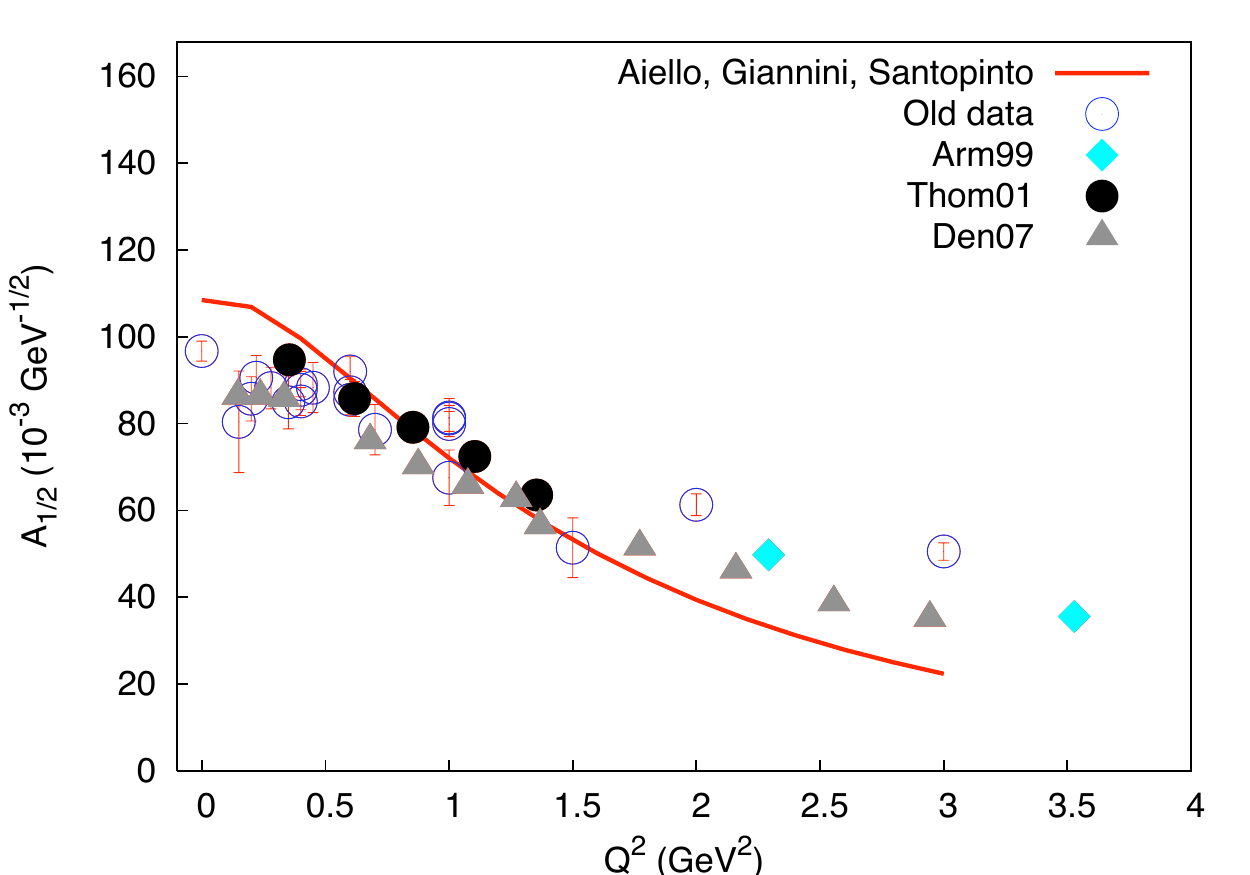}
  \caption{(Color online) The transverse helicity amplitude for the S11(1525) resonance, calculated with the hCQM of Eq. (\ref{hCQM}). Data are from \cite{bu,arm99,thom01,den07}. The photon point is also shown \cite{pdg}.}
\end{figure}

The hCQM seems to provide realistic three-quark wave functions and the main reason is the presence of the hypercoulomb term. In fact, using the analytical version of the model \cite{sig}, the wave functions in lower order coincide with the ones given by the hypercoulomb potential of Eq. (\ref{hcb}), nevertheless the results for the helicity amplitudes are very similar to the ones predicted by the hCQM.

\section{The elastic nucleon form factors}

Recently at Jlab data \cite{rap_exp}  the ratio
\begin{equation}
R_p~=~\mu_p \frac {G^p_E(Q^2)}{G^p_M(Q^2)}
\end{equation}
has been obtained directly from the polarization asymmetry measured in the elastic scattering of polarized electrons on polarized protons. This ratio
deviates strongly from zero with increasing values of $Q^2$ and seems to tend to vanish at high  $Q^2$. The problem arises of discrepancies with respect to data obtained with the Rosenbluth plot, however the Jlab data have triggered a renewed interest in the problem of describing the nucleon elastic form factors within quark models. In particular, the attention is focused on the possible zero of the electric form factor at high $Q^2$. It should be mentioned that extrapolating at high $Q^2$ the fit proposed many years ago \cite{IJL}, one gets a strong depletion of the ratio $R_p$ and a zero not far from 10 $ GeV^2$. Calculations based on microscopic model for the nucleon structure provide a decreasing behavior for $R_p$, in some cases giving rise also to a zero  \cite{ff_altri}.

\begin{figure}[h]
  \includegraphics[height=.4\textheight]{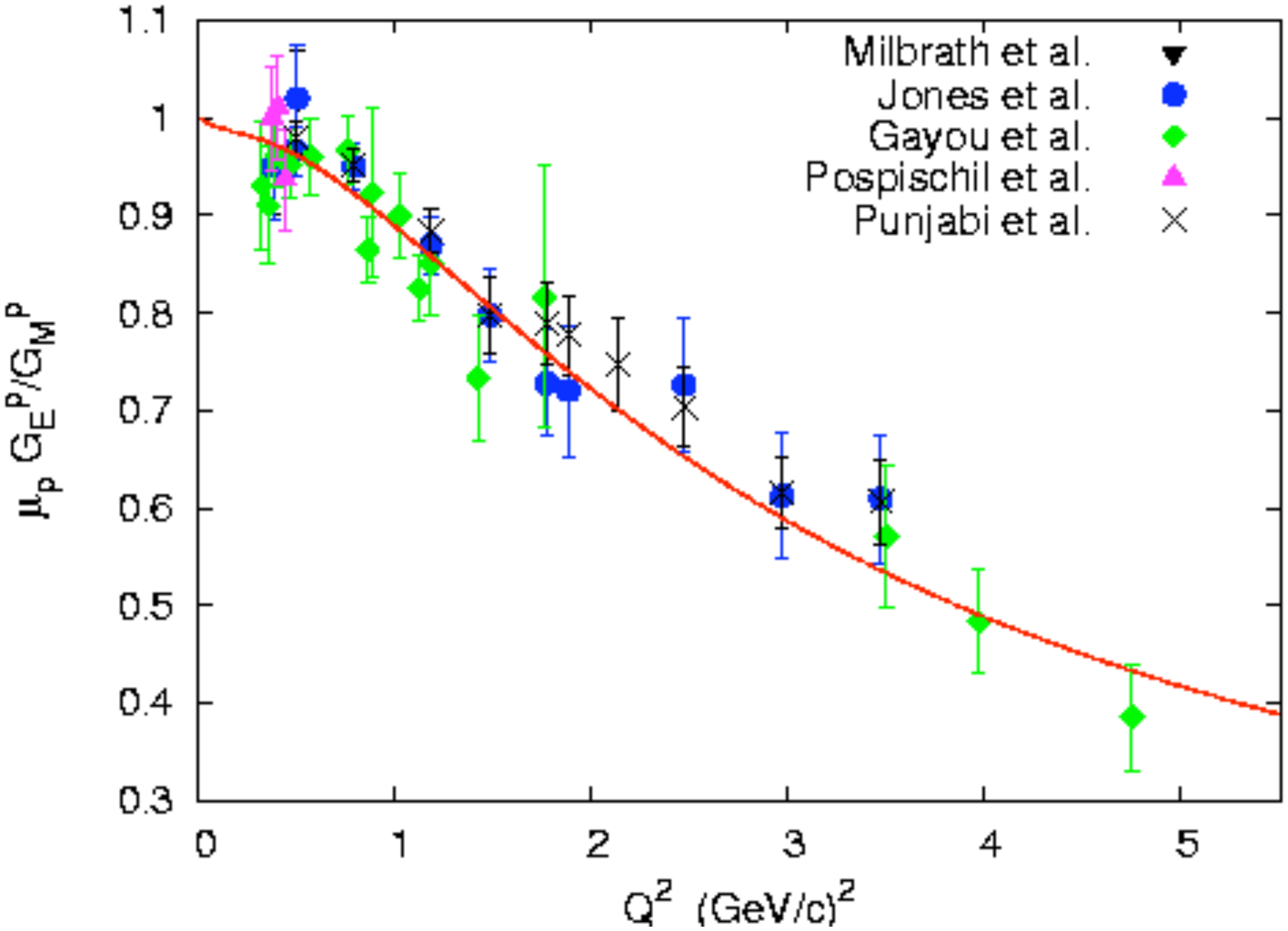}
  \caption{(Color online) The ratio $\mu_p G_E^p/G_M^p$ from
polarization transfer compared with the relativistic hCQM calculation with
      constituent quark form factors (solid line).The experimental data are taken from~\cite{rap_exp}. The Figure is taken from \cite{ff_07} (APS Copyright).}
\end{figure}

The hCQM provides nucleon wave functions which can be used for the calculation of the elastic form factor. However, since the calculated proton radius turns out to be about $0.5 fm$, the resulting form factors are not good and one can expect that relativity is needed. In fact, we have shown that the depletion of $R_p$ is a relativistic effect \cite{rap}: by simply boosting \cite{mds} the nucleon states to the Breit system one gets a deviation from 1 which reaches the value of about $15 \%$ at $Q^2=2 GeV^2$.

\begin{figure}[h]
  \includegraphics[height=.4\textheight]{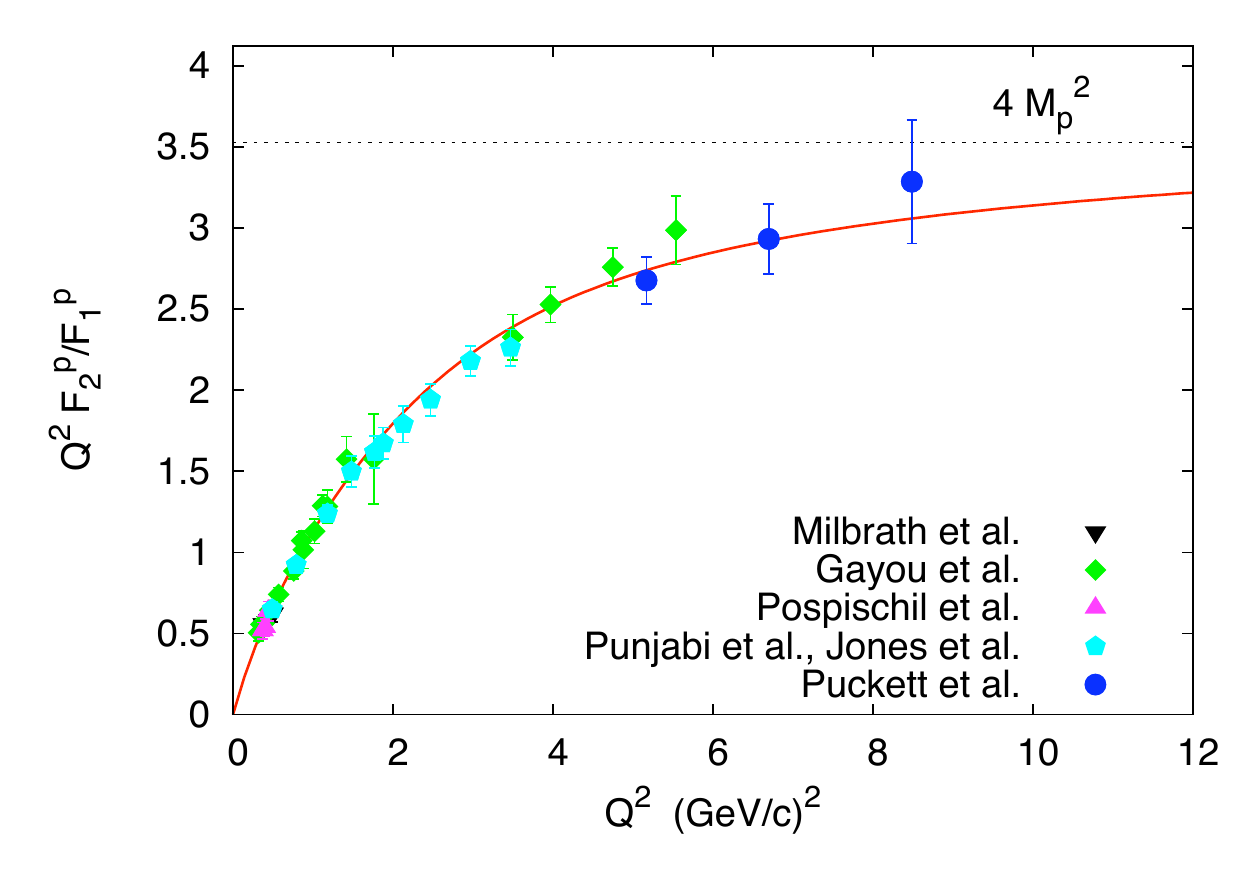}
  \caption{(Color online) The ratio $\mu_p G_E^p/G_M^p$ from
polarization transfer compared with the relativistic hCQM calculation with
      constituent quark form factors (solid line).The experimental data are taken from~\cite{rap_exp,Puckett}. The Figure is taken from \cite{ff_10} (APS Copyright).}
\end{figure}

For this reason, the hCQM has been fully relativized \cite{ff_07} using the Dirac Relativistic dynamics in the point form (PF). The resulting predicted form factors are in good agreement, specially the magnetic ones, however, the ratio tends to a constant value of about 0.6, with a behavior different from the observed one. 

In order to understand what is missing, one can remind that CQs are effective degrees of freedom, which can acquire mass and even size. The latter statement is supported by the 
recent analysis of the deep inelastic electron-proton scattering  \cite{ricco}, in which there is evidence of quarks having a finite size. One can therefore think  that CQ form factors must have a role also in the description of the nucleon from factors.

In Fig. 3 we report the relativistic hCQM result taking into account quark form factors. They have determined in order to reproduce at the same time the ratio $R_p$ and the proton magnetic, neutron electric and magnetic form factors up to $Q^2=5 GeV^2$, with very nice results \cite{ff_07}. 

The data seem to have a linearly decreasing behavior and the question is if they reach zero or not. A further information is given by very recent data \cite{Puckett}, which extended the measure of $R_p$ up to $8.5 GeV^2$. For this reason we have extended the calculation with the hCQM \cite{ff_07} up to $12 GeV^2$ \cite{ff_10}, without modifying any parameter. In order to study the high $Q^2$ behavior of $R_p$ one can alternatively consider the ratio
\begin{equation}
F_p~=~ Q^2\frac {F^p_2(Q^2)}{F^p_1(Q^2)}
\end{equation}
The zero of $R_p$ occurs when $F_p = 4 M_p^2$, $M_p$ being the proton mass. In Fig. 4 we report the predicted high $Q^2$ behavior in comparison with the new data. While the first two new data are in agreement with the hCQM prediction, which does not tend towards a zero value of $R_p$, the last point is compatible with a dip in the electric form factor. The situation will be hopefully clarified by the future experiments to be performed at Jlab.

\section{Conclusion}

The hCQM seems to provide realistic quark wave functions even in its non relativistic formulation. Its predictions give rise to an overall agreement with the observed helicity amplitudes, specially in the medium $Q^2$ region, where the quark degrees of freedom are expected to dominate. At low $Q^2$ there is a lack of strength, which can be attributed to the missing quark-antiquark effects. The relativization of the model is certainly an important issue, although it appears to give relevant contributions in the case of the elastic form factor, while the helicity amplitudes are only slightly affected. 
The missing dynamical mechanism is given by the quark-antiquark creation, pointing towards the unquenching of the CQMs. Important progress in this direction is provided by recent work, in particular for the baryons  by \cite{UCQ,es_coc}, while for the mesons by  \cite{UCQmeson}. Such unquenched CQs will allow a consistent treatment of the baryon spectrum, the elastic nucleon properties and the electroproduction of mesons.

\end{document}